\documentstyle[epsfig]{aipproc}

\def\ltsima{$\; \buildrel < \over \sim \;$}
\def\simlt{\lower.5ex\hbox{\ltsima}} % < over ~
\def\gtsima{$\; \buildrel > \over \sim \;$}
\def\simgt{\lower.5ex\hbox{\gtsima}} % > over ~

\def\V{\hbox{$\rm V$}}

\newcommand{\etal} {{\it et~al.\ }}
\begin{document}
%\input{psfig.tex}
 
%\title{The Fading Counterpart of GRB~970228, Six Months Later}
%\title{Deep Imaging of the GRB~970228 Field with HST-STIS}
\title{HST/STIS Observations of the Optical Counterpart to GRB~970228}

\author{Andrew S. Fruchter$^1$,
Elena Pian$^2$,
Stephen E. Thorsett$^3$,
%Henry Ferguson\altaffilmark{1},
Rosa Gonzalez$^1$,
Kailash C. Sahu$^1$,
Max Mutchler$^1$,
Filippo Frontera$^{2,6}$,
Titus Galama$^7$,
Paul Groot$^7$,
Richard Hook$^9$
Chryssa Kouveliotou$^8$,
Mario Livio$^1$,
Duccio Macchetto$^{1}$,
Jan van Paradijs$^{7}$,
Eliana Palazzi$^{2}$,
Larry Petro$^1$, 
Marco Tavani$^{4,5}$}

\address{$^{1}$Space Telescope Science Institute, 3700 San Martin 
Drive, Baltimore, MD 21218, USA\\
$^{2}$Istituto di Tecnologie e Studio delle Radiazioni 
Extraterrestri, C.N.R., Via Gobetti 101, I-40129 Bologna, Italy \\
$^{3}$Joseph Henry Laboratories and Dept.\ of Physics, Princeton 
University, Princeton, NJ 08544, USA\\
$^{4}$Columbia Astrophysics Laboratory, Columbia 
University, New York, NY 10027, USA \\
$^{5}$Istituto di Fisica Cosmica e Tecnologie Relative, 
C.N.R., Via Bassini 15, I-20133 Milano, Italy\\
$^{6}$Dip. Fisica, Universit\`a di Ferrara, Via Paradiso 
12, I-44100 Ferrara, Italy\\
$^{7}$Astronomical Institute ``Antonton Pannekoek'', University
of Amsterdam, Kruislaan 403, 1098 SJ Amsterdam, The Netherlands\\
$^8$NASA Marshall Space Flight Center, ES-84, Huntsville, AL 35812, USA\\
$^9$Space Telescope European Coordinating Facility, D-85748 Garching, Germany\\
}

\maketitle

\begin{abstract}
We report on observations of the fading optical counterpart of the
gamma-ray burst GRB~970228, made on 4~September~1997 using the STIS
CCD on the Hubble Space Telescope. The unresolved counterpart
is detected at $V=28.0\pm0.25$, consistent with a continued power-law
decline with exponent $-1.14 \pm 0.05$.  No proper motion is detected, in
contradiction of an earlier claim. The counterpart is located
within, but near the edge of, a faint extended source with 
diameter $\sim 0\farcs8$ and integrated magnitude
$V=25.7\pm0.25$ . Comparison with WFPC2 data taken one month after
the initial burst and NTT data taken on March 13
shows no evidence for fading of the extended emission.

%   The counts and colors of background galaxies in the
%   WFPC2 images as well as the 100 micron IRAS map 
%   (Schlegel, Finkbeiner and Davis 1997)
%   imply a Galactic extinction in the direction of GRB~970228
%   of $A_v = 0.7 \pm 0.15$. 
After adjusting for the probable Galactic extinction in the direction
of GRB~970228 of $A_v \sim 0.7$,
we find that the observed nebula is consistent with the sizes of
galaxies of comparable magnitude found in the Hubble Deep Field and
other deep HST images, and that only 2\% of the sky is covered
by galaxies of comparable magnitude and similar or greater surface brightness.
We therefore conclude that the extended source observed about GRB~970228
is most likely a galaxy at moderate redshift, and is almost 
certainly the host of the gamma-ray burst.

\end{abstract}
%\keywords{Cosmology: observations --- galaxies: starburst --- 
%gamma rays: bursts --- stars: formation}

\section*{Introduction}

Identification and analysis of long wavelength counterparts of
gamma-ray bursts (GRBs) has for many years been considered a promising path
towards understanding the nature of the burst events.  But while
many attempts
have been made to identify GRB counterparts 
(Schaefer \etal 1987, Schaeffer 1992, Fenimore \etal 1993, Larson 1997) 
until recently
the uncertainties in the position of the gamma-ray sources proved too
large to allow sufficiently sensitive surveys for associated optical
transients (OTs). The situation improved dramatically in early 1997, when
gamma- and X-ray observations by the BeppoSAX satellite provided a
sub-arcminute position of burst GRB~970228, allowing the first firm
optical identification of a fading GRB counterpart (van Paradijs \etal 1997).  
A
second optical GRB counterpart, this time of GRB~970508, was
discovered two months later (Bond 1997, Djorgovski \etal 1997).  

Although broadly similar in their fading behavior, phenomenological
differences between the two counterparts in the weeks after their
discoveries seemed to compound rather than clarify the mystery of the
GRBs.  HST imagery of GRB~970228 suggested the presence of a
nebulosity centered $0\farcs3$~arcsecond from the point-like fading transient
source (Sahu \etal 1997), while no extended source brighter than $R =
24.5$ has been found near GRB~970508 (Pian \etal 1997).
%The light curves of the two optical
%transients, monitored for as long as possible from ground,
%have shown for GRB970508 a power-law decrease of the flux 
%till one month later than discovery (Pedersen...; 
%Castro-Tirado...; Pian et al. 1997, and references therein).  
%On the other hand, the optical trend of GRB~970228 in a
%comparable time interval, decomposed from the extended
%source, shows a flattening and is not well described by
%a single power-law (Galama et al. 1997).
Furthermore, a proper motion of 550~mas~yr$^{-1}$ was
reported (Caraveo \etal 1997) for GRB~970228, 
though the measurement was disputed (Sahu \etal 1997b).
Additionally, tentative evidence for the fading of the adjacent
nebulosity (Metzger \etal 1997a) was proposed. Either result would
ineluctably lead to the conclusion that GRB~970228 was a Galactic
event. In contrast, the measurement of absorption lines with redshift
$z\ge0.835$ in the spectrum of GRB~970508 (Metzger \etal 1997b) demonstrates
its extragalactic nature.   
 
To help resolve the situation, we have reobserved the GRB~970228 with HST 
six months after the initial outburst.  
Although the earlier HST observations of GRB~970228 employed WFPC2,
we have availed ourselves of the newly installed STIS CCD camera.
The excellent
throughput and broad bandpass of this instrument,  combined
with the long time baseline since the gamma-ray burst, 
provide us with a  superb opportunity
to study the nature of the source and its environment. 

%In 
%this paper we present the reduction and analysis of these
%new observations, as well as implications for the nature
%of gamma-ray bursts and their hosts.
%benefit from lower background
%noise levels and are deeper than previous WFPC images.  
%Our data
%reduction and analysis procedures are described in \S2, our results
%are presented and compared with previous work in \S3, and our
%conclusions about the nature of the point-like and nebular
%counterparts are presented in \S4.

\section*{Observations, Image Reduction and Photometry}

The field of GRB~970228 was imaged during two HST orbits
on 1997~September~4
from 15:50:33 to 18:22:41 UT, using the
STIS CCD in Clear Aperture (50CCD) mode.  Two exposures
of 575s each were taken at each of four dither positions
for a total exposure time of 4600s.  The exposures were dithered
to allow removal of hot pixels and to obtain
the highest possible resolution.  The images were bias and
dark subtracted,
and flat-fielded using the STIS
pipeline.  The final image was created and cleaned of cosmic rays
and hot pixels
using the Drizzle and Blot algorithms developed for the
Hubble Deep Field (HDF) (Williams \etal 1996, Fruchter and Hook 1997).
An output pixel size of  $0\farcs025$
across, or one-half the size of the input pixels, was used.
%and  a ``pixfrac" of 1.0 were used.
%In order to make the faint extended nebula more easily visible,
%we also block-averaged the high resolution image onto pixels
%$0\farcs1$ across.  Both  of these images are shown in
%Figure 1.

\begin{figure} % fig 1
\centerline{\epsfig{file=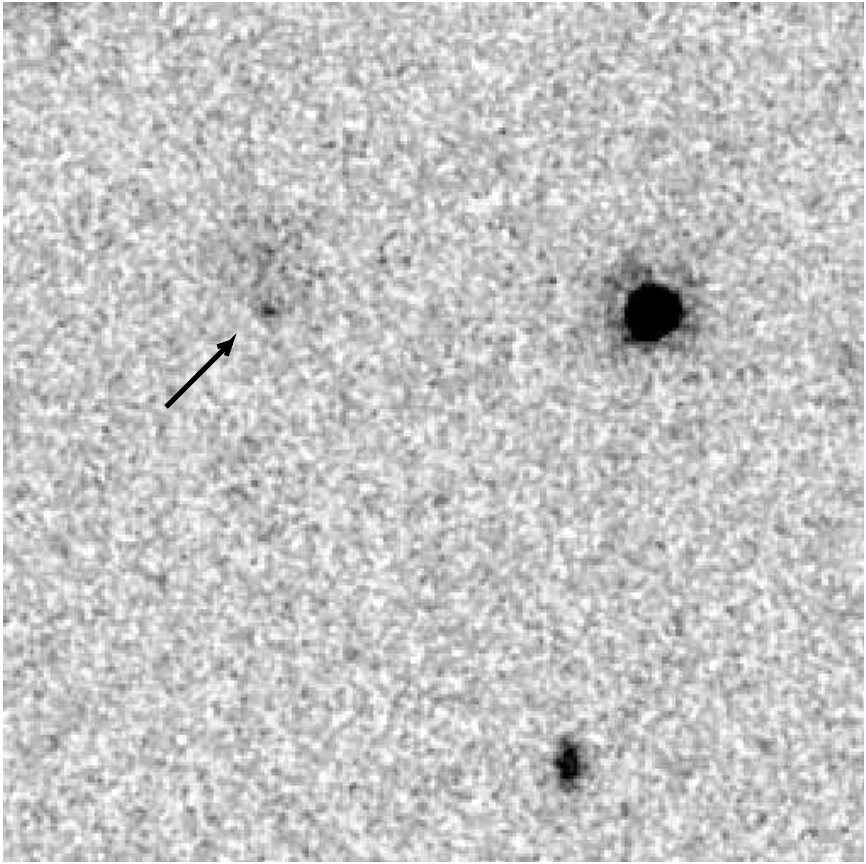,height=4.5in,width=4.5in}}
\vspace{10pt}
\caption{The STIS image of GRB~970228.  North is up; East
is to the left.  An arrow points
to the OT.  The nebula can be seen
extending to the north of the OT.}
\label{fig1}
\end{figure}

%The photometric calibration of the images was done using the synthetic
%photometry package SYNPHOT in IRAF/STSDAS,  although total throughput
%was renormalized by approximately 12\% to agree with the on-orbit 
%recalibration of the STIS CCD by
%Landsman (1997).  
%This adjustment
%agrees well both with estimates derived by
%comparison of the stellar magnitudes
%in the WFPC2 F606W image of the GRB~970228 field with
%those of the STIS image, and with a separate calibration of the STIS CCD
%done by Sahu \cite{S97}.  

%The STIS CCD in clear mode has a broad bandpass, with
%a significant response from
%200 to 900 nm that peaks near 600nm.  As a result,
%STIS instrumental magnitudes are most accurately
%translated into the standard filter set by quoting
%the result as a $\V$ magnitude, but in any case
%knowledge of an object's intrinsic spectrum is required 
%for an accurate conversion.   Here
% we are fortunate to have an earlier
%image of the field in both $\V$ and $\I$ with the
%WFPC2.  We therefore can use colors determined
%in these earlier observations to interpret the
%STIS data.  However, to the extent that the WFPC2 colors are
%in error (either due to low signal-to-noise or an
%actual change in the color of the object over time)
%our derived magnitudes from the STIS data will be
%biased by about 0.5 magnitudes for each magnitude
%of error in $\V - \I$ \cite{Sahu}.

The magnitude of the OT was determined
from the drizzle image via aperture photometry.  The flux
in an aperture of
radius four drizzled pixels, or $0\farcs$, was determined, and
our best estimate of the surrounding nebular background was
subtracted.  An aperture correction of $0.50$ magnitudes
was derived for this aperture using the bright star
visible in Figure 1 to the west of the nebula.
%image (the STIS shows signficant PSF variability over the
%field-of -view; therefore we preferred  chose to use a single
%nearby star for the template
%rather than the average of a larger number of more
%distant objects).    
We find a total magnitude for the point source OT of 
$\V = 28.0 \pm 0.25$. 

\begin{figure}[t!] % fig 2
\centerline{\epsfig{file=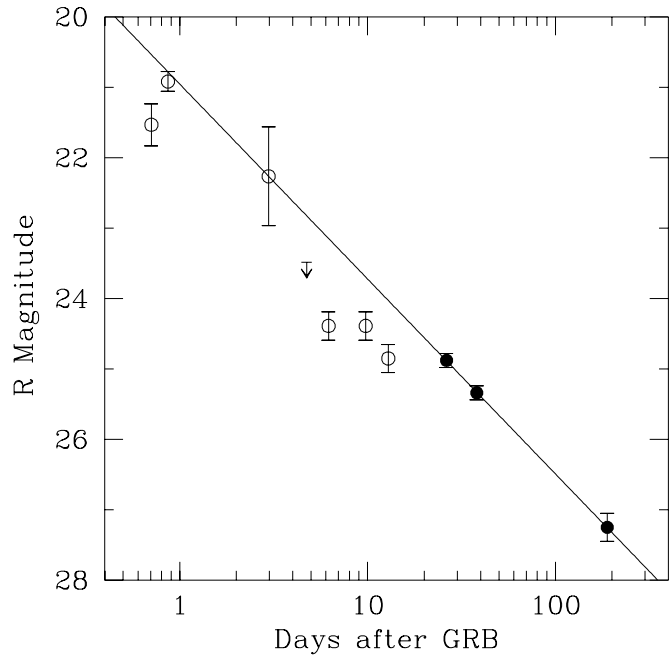,height=4.5in,width=4.5in}}
\vspace{10pt}
\caption{The R magnitude of the OT as a function of
time.  A nebular R magnitude of 25.3 has been subtracted from all
non-HST magnitudes.  The line shows the best fit power law
through the three HST observations (shown in black).  See
Rees and Meszaros (1997) for a possible explanation for the deviation of
the observed points deviation from the power law.}
\label{fig2}
\end{figure}

In figure~2 we plot the magnitude of the OT as a function of time
since the burst last February.  The STIS magnitude has been converted
to R by interpolating the WFPC2 V and I colors.  
A power law
of the form $f(t) = a_0*t^{\alpha}$ has been fitted to the
HST points and extrapolated back to earlier times.  We find a best fit of
$\alpha = -1.14 \pm 0.05$.
All non-HST photometry has been adjusted under the assumption
that the R magnitude of the nebula is 25.3, which was
obtained by interpolating
the STSIS V and WFPC2 I measurements (see below).

We determined the magnitude of the nebula
by summing all pixels in a region of approximately
$1.4$ sq. arcsec. surrounding the object.   The flux of the
point source was then subtacted
from the sum.   We derive a magnitude of $\V = 25.7 \pm 0.25$ for
the nebula.
Due to the very wide bandpass of the STIS clear observations
(the only constraint on the bandpass comes from the optics
and the response of the CCD),
the primary source of photometric error
is the correction of the STIS magnitude
for the color of the object.   This is particularly large
for the nebula, as the only measurement of the color of
this object comes from the previous, rather noisy,
Planetary Camera observations of
the field (these images are discussed in more detail in the next section).

In order to determine whether the optical transient
displayed proper motion or the nebula faded we have
compared the STIS images to the previous HST 
WFPC2 images obtained obtained of GRB~970228 (Sahu \etal 1997a).
The images taken on March~26 with WFPC2 
provide a baseline of 162 days and are used here
to look for proper motion of the optical transient.
%The WFPC2 images were reduced as describe in Sahu \etal (1997b). 
%The four reference stars used in that work as positional anchors
%for their previous proper motion study
%are also visible in the STIS image.
%The centroids of the four reference stars used in Sahu \etal (1997b) 
%were determined using 2-dimensional
%Gaussian fits to both the WFPC2 and STIS images. 
%The positional accuracy for the reference stars is 
%typically  2 to 3 mas in each coordinate at each epoch, 
%while for the relatively faint GRB it is $\sim$5 mas.  
%The measured pixel-coordinates were first corrected for the camera's geometric 
%distortion using the Gilmozzi \etal\ (1995) solutions  for the WFPC2 images,
%and the solution derived by Malmuth et al. (1997, STIS Calibration Report) 
%for the STIS images.  
%The STIS pixel-coordinates were then transformed to the corresponding
%WFPC2 pixel-coordinates. This was done taking the rotation, translation and the 
%image scale into account in a single step using the 4~reference stars 
%used in Sahu \etal 1997b, 
%and assuming zero mean motion. The same procedure was repeated for 
%the WFPC2 V and I-band images separately.

The positions of the four reference stars used in Sahu \etal (1997b)
agree with their positions in the STIS images  to within the expected 
uncertainties of 2 to 3 mas, 
which shows that the transformations between the two images have
been done correctly.   The uncertainty in the position of the OT
is about 10 mas in each of the two colors.
We find that any motion of the GRB between the two epochs
is less than about 16
mas. 
This corresponds to a motion of less than 36 mas
per year. This is a factor of $\sim 15$ less
than the value claimed by Caraveo \etal
(1997), and improves the upper limit on the proper motion
reported by Sahu \etal (1997b) by a factor of six.

To check on the photometry of the optical transient and nebula,
the point
source magnitude was determined by using 
circular apertures of radii 1 and 3 pixels in the WFPC2 images and adjusting
the observed fluxes according to the aperture
corrections found by Holtzman \etal (1995).
The nebular magnitude was redetermined in the WFPC2 images
by taking the sum
of all counts above sky in a box approximately $1\farcs5 \times 1\farcs0$, and
subtracting the counts (estimated as above) attributable to
the point source.    The position of this box was determined
by the position of the nebula in the STIS image.  It is, however,
somewhat larger than the observed nebula in all directions.
Averaging together the two WFPC2 observations, we obtain an
I  magnitude of $24.4 \pm 0.2$
and a V magnitude of $25.5 \pm 0.2$.  
These magnitudes are easily consistent with
the STIS observation. 

We have also re-examined the NTT observation of March 13 (Galama \etal 1997) to
further test whether the nebular magnitude may have varied with time.
We have again used the stellar image $\sim 2\farcs5$
to the west of the OT as a point spread function.    We find that
we can subtract a point source from the position of the OT which is
fainter than the extrapolation of the power-law, yet which leaves
behind a ``neblula'' which is as faint, or fainter than, the HST
nebular magnitude.  Thus, we find no evidence
that the nebula has changed magnitude
with time. 

\section*{Astrophysical Implications}

There is little room for doubt that the fading point source is
associated with the gamma-ray event.  Between 28 February and 4
September 1997, the source faded by a factor
of about 350, and as shown in Figure 2, this dramatic fall
in luminosity largely followed a power-law whose index, $-1.14 \pm 0.05$,
 is within the errors
indistinguishable from the index of
power-law decline of the optical counterpart to GRB~970508 (Pian \etal 1997).
Given the lack of any other astrophysical objects with similar behavior,
and the theoretical prediction of a power-law fall-off with time
of the luminosity of afterglow (Meszaros and Rees 1997), we believe there
is no reasonable alternative to the conclusion
 that we are observing the optical afterglow
of GRB~970228. 

Furthermore, in
simple blast wave models, a break in the power-law
 to $F\sim t^{-1.8}$ is expected
(Wijers, Rees and Meszaros 1997) when the remnant enters a Sedov-Taylor phase after
sweeping up a rest mass energy equal to its initial energy $E$ at time:
\begin{equation}
t\approx 1\mbox{\,yr}\left(\frac{E_{52}}{n}\right)^{1/3},
\end{equation}
where $n$ is the density of the surrounding medium in protons
per cubic centimeter.   However, were
the GRB a Galactic rather than an extragalactic phenomenon, the amount
of energy available would only be of order $10^{41}$ ergs, and for
any imaginable density the break would occur on a timescale of days
rather than many months.  Therefore, the power law fit is in itself a strong
argument for the extragalactic nature of the burst. 

If the burst is extragalactic, then it is natural to inquire whether
the apparently constant nebula seen under the OT is the host galaxy.
The Galactic extinction in the direction of GRB~970228  has been estimated
as $A_v \sim 0.7$ (Burstein and Heiles 1982, Schlegel, Finkbeiner and
Davis 1997) -- a figure that we have been able
to independently verify by comparing the 
counts and colors of background galaxies
in the WFPC2 field with the HDF.   
Adjusting the surface brightness limit
to reflect the $\sim 0.7$ mags of extinction in the
direction of GRB~970228,  we find that only about 2\% of the
sky in the HDF is covered by galaxies
of comparable magnitude, and that the size of the putative host of GRB~970228,
while larger than the mean 25th magnitude galaxy in the HDF, is not,
by any means, extraordinary. 

Although we have no spectroscopic information on the redshift of this object
nor do we have sufficient colors to attempt a photometric redshift (though
planned NICMOS observations may rectify this problem),
we can attempt to place a crude constraint on the plausible redshift simply from
the luminosity of the object.  Were the object closer than $z \sim 0.5$
it would be more than four magnitudes fainter than $L_*$ (Lilly \etal 1995),
and this is unlikely even given the steep luminosity function at
that redshift (Ellis \etal 1996).  On the other hand, the apparent
host is as bright as any ``U dropout'' in the HDF (Madau \etal 1996), 
and therefore
would be an unusually bright galaxy were it at the typical
redshift of these dropouts, $z \sim 2.5$.    Thus a
plausible redshift range for the host is $0.5 \simlt z \simlt 2.5$.
However, while the luminosity function of galaxies is a rather blunt instrument
for estimating the redshifts of GRB hosts, we will show in the
journal paper associated with this work that GRB hosts may prove a
rather better tool for determining the luminosity function of 
galaxies.

We thank the Director of STScI, Bob Williams,
for allocating Director's Discretionary time to
this program.

\end{document}